\documentclass[prl,twocolumn,showpacs,superscriptaddress]{revtex4}
\usepackage{graphicx}
\usepackage{amssymb}
\usepackage{amsmath}
\usepackage{amsfonts}
\usepackage{mathrsfs}
\begin{document}
\title{DNA bubble dynamics as a quantum Coulomb problem}
\author{Hans C. Fogedby}
\email{fogedby@phys.au.dk} \affiliation{Department of Physics and
Astronomy, University of Aarhus, DK-8000, Aarhus C, Denmark}
\affiliation{Niels Bohr Institute for Astronomy, Physics, and
Geophysics, Blegdamsvej 17, DK-2100, Copenhagen {\O}, Denmark}
\affiliation{NORDITA, Blegdamsvej 17, DK-2100, Copenhagen {\O},
Denmark}
\author{Ralf Metzler}
\email{metz@nordita.dk}
\affiliation{NORDITA, Blegdamsvej 17, DK-2100, Copenhagen {\O}, Denmark}
\affiliation{Department of Physics, University of Ottawa, 150 Louis Pasteur,
Ottawa, Ontario  K1N 6N5, Canada}
%\date{\today}
\begin{abstract}
We study the dynamics of denaturation bubbles in double-stranded DNA
on the basis of the Poland-Scheraga model. We demonstrate that the
associated Fokker-Planck equation is equivalent to a Coulomb
problem. Below the melting temperature the bubble lifetime is
associated with the continuum of scattering states of the repulsive
Coulomb potential, at the melting temperature the Coulomb potential
vanishes and the underlying first exit dynamics exhibits a long time
power law tail, above the melting temperature, corresponding to an
attractive Coulomb potential, the long time dynamics is controlled
by the lowest bound state. Correlations and finite size effects are
discussed.
\end{abstract}
\maketitle

\emph{Introduction.}
The dynamics of bubble formation in double-stranded DNA (dsDNA) is a problem
of high current interest in biological and statistical physics. Under
physiological conditions the Watson-Crick double helix is the equilibrium
structure, its stability effected by hydrogen-bonding of base-pairs and
stacking between nearest neighbor pairs of base-pairs
\cite{Kornberg74,Watson53}. By variation of temperature or pH-value
double-stranded DNA progressively denatures, yielding regions of
single-stranded DNA
(ssDNA), until the double-strand is fully denatured. This is the
helix-coil transition associated with the melting temperature
$T_{\text{m}}$ \cite{Poland70}.

Subject to thermal fluctuations dsDNA spontaneously unzips and
forms flexible single-stranded DNA bubbles ranging in size from a
few to some hundred broken base-pairs, depending on $T$ and salt
conditions \cite{Poland70,Gueron87,Krueger06}. Assuming that the
bubble-forming dynamics takes place on a slower time scale than
the equilibration of the ssDNA strands constituting the bubbles,
this DNA-breathing can be interpreted as a random walk in the 1D
coordinate $x$, the number of denatured base-pairs.

On the basis of the Poland-Scheraga model for DNA-melting \cite{Poland66}
DNA-breathing has been studied in terms of a continuous Fokker-Planck
equation \cite{Hwa01,Hanke03}, and of a discrete master equation and
stochastic simulation \cite{Banik05,Metzler04}. In the present Letter
we show that the Fokker-Planck equation for bubble breathing is equivalent
to a quantum Coulomb problem with a repulsive
potential above $T_{\text{m}}$ and an attractive potential below
$T_{\text{m}}$. This mapping allows us to discuss DNA bubble
dynamics in terms of the spectrum of a 'hydrogen-like' system and to
derive several exact results such as the exact scaling of the bubble
survival behavior and the associated correlations.

\emph{Static and dynamic model.} The Poland-Scheraga free energy
for the bubble statistics has the form
\cite{Poland70,Hanke03,Krueger06}
\begin{eqnarray}
\label{free}
\mathscr{F}=\gamma_0+\gamma x+c\ln x,
\end{eqnarray}
where $x\ge 0$ is the bubble size in units of base pairs. We here
assume a continuum formulation and imply a cutoff for $x\sim 1$.
$\gamma_0$ is the free energy barrier for initial bubble
formation, $\gamma x$ is the free energy for the dissociation of
$x$ base pairs, and $c\ln x$ the entropy loss factor associated
with the formation of a closed polymer ring. The free energy
density $\gamma=\gamma_1(1-T/T_{\text{m}})$, where $T_{\text{m}}$
is the melting temperature. From experimental data one extracts
approximate values for the parameters. In units of $kT_{\text{r}}$
with reference temperature $T_{\text{r}}=37^{\circ}$C, we have
$\gamma_0\approx10 kT_{\text{r}}$, $\gamma_1\approx 4 kT_{\text{r}}$,
and $c\approx 2 kT_{\text{r}}$; the melting temperature for standard
salt conditions is in the range $T_{\text{m}}\approx 70-100^{\circ}$C,
depending on the relative content of weaker AT and stronger GC
Watson-Crick base-pairs.

The stochastic bubble dynamics is governed by the Langevin equation
with Gaussian white noise $\xi(t)$,
\begin{equation}
\label{lan}
\frac{dx}{dt}=-D\frac{d\mathscr{F}}{dx}+\xi,
\quad\langle\xi\xi\rangle(t)=2DkT\delta(t),
\end{equation}
where the kinetic coefficient $D$ of dimension $(kT_{\text{r}})^{-1}
\mathrm{s}^{-1}$ sets the overall time scale of the dynamics:
$[DkT_{\text{r}}]^{-1}\sim\mu$s. With dimensionless parameters
$\mu=c/2kT$ and $\epsilon=(\gamma_1/2kT)(T/T_{\text{m}}-1)$, and
measuring time in units of $\mu s$, the Fokker-Planck equation corresponding
to (\ref{lan}) reads
\begin{eqnarray}
\frac{\partial P}{\partial t}=\frac{\partial}{\partial
x}\left(\frac{\mu}{x}-\epsilon\right)P+
\frac{1}{2}\frac{\partial^2P}{\partial x^2}.\label{fokker}
\end{eqnarray}
Note that close to the physiological temperature $T_{\mathrm{r}}$,
$\mu\approx 1$ and $\epsilon\approx2(T/T_{\text{m}}-1)$.

\emph{General results.} Eliminating the first order term by means of the
substitution $P=e^{\epsilon x}x^{-\mu}\tilde P$, $\tilde P$
satisfies
\begin{equation}
\label{schroedinger}
-\frac{\partial\tilde P}{\partial t}=
-\frac{1}{2}\frac{\partial^2\tilde P}{\partial x^2}+
\left(\frac{\mu(\mu+1)}{2x^2}-
\frac{\mu\epsilon}{x}+\frac{\epsilon^2}{2}\right)\tilde P.
\end{equation}
This is the imaginary time Schr\"{o}dinger equation for a particle with unit
mass in the potential
$V(x)=\mu(\mu+1)/2x^2-\mu\epsilon/x+\epsilon^2/2$, i.e., subject
to the centrifugal barrier $\mu(\mu+1)/x^2$ for an orbital state
with angular momentum $\mu$ and Coulomb potential
$-\mu\epsilon/x$. Introducing the Hamiltonian
$H=-(1/2)d^2/dx^2+V(x)$ and expanding $\tilde P$ on the normalized
eigenstates $\Psi_n$, $H\Psi_n=E_n\Psi_n$, the
transition probability $P(x,x_0,t)$ from initial bubble size
$x_0$ to a final bubble size $x$ at time $t$ yields in closed form,
\begin{eqnarray}
P(x,x_0,t)=e^{\epsilon(x-x_0)}\left(\frac{x}{x_0}\right)^{-\mu}
\sum_ne^{-E_nt}\Psi_n(x)\Psi_n(x_0). \label{prob}
\end{eqnarray}
Here the completeness of $\Psi_n$ ensures the initial condition
$P(x,x_0,0)=\delta(x-x_0)$. Moreover, in order to account for the
absorbing boundary condition for vanishing bubble size we choose
$\Psi_n(0)=0$. We also note that for a finite strand of length $L$,
i.e., a maximum bubble size of $L$, we have in addition the
absorbing condition $\Psi_n(L)=0$ for complete denaturation. Expression
(\ref{prob}) is the basis for our discussion of DNA-breathing,
relating the dynamics to the spectrum of
eigenstates, i.e., the bound and scattering states of the
corresponding Coulomb problem \cite{Landau59c}.

\begin{figure}
\includegraphics[width=.64\hsize]{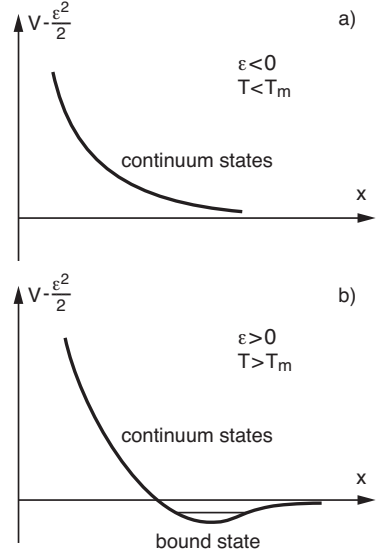}
\caption{Schematic of the potential $V(x)-\epsilon^2/2$ of the Schr{\"o}dinger
Eq.~(\ref{schroedinger}). a) $T<T_{\text{m}}$: The potential is repulsive,
yielding a continuous spectrum. The bubble fluctuations correspond to
a Brownian walk process in bubble size $x$ before collapse at $x=0$.
b) $T<T_{\text{m}}$. The potential is attractive and can trap a series of
bound states. At long times the lowest bound state indicated in the figure
controls the behavior. The bubbles increase in size leading to denaturation.
} \label{fig1}
\end{figure}

The transition probability $P$ is controlled by the Coulomb
spectrum. Below the melting temperature $T_{\text{m}}$
($\epsilon\propto(T/T_{\text{m}}-1)<0$), the Coulomb problem is
repulsive and the states form a continuum, corresponding to a random
walk in bubble size terminating in bubble closure $(x=0)$. At the
melting temperature ($\epsilon=0$), the Coulomb potential is absent
and the continuum of states is governed by the centrifugal barrier
alone, including the limiting case of a regular random walk. Above
the melting temperature ($\epsilon>0$), the Coulomb potential is
attractive and can trap an infinity of bound states; at long times
the lowest bound state dominates the bubble dynamics, corresponding
to denaturation of the DNA chain. In Fig.~\ref{fig1} we depict the
potential in the two cases $\epsilon \gtrless 0$.

\emph{(i) Long times for $T\le T_{\text{m}}$.} At long times and
fixed $x$ and $x_0$, it follows from (\ref{prob}) that the
transition probability is controlled by the bottom of the energy
spectrum. Below and at $T_{\text{m}}$ the spectrum is continuous
with lower bound $\epsilon^2/2$. Setting $E_k=\epsilon^2/2+k^2/2$ in
terms of the wave number $k$ and noting that
$\Psi_k(x)\propto(kx)^{1+\mu}$ for small $kx$ we have
$P\propto\exp(-|\epsilon|(x-x_0)) (x/x_0)^{-\mu}
\exp(-\epsilon^2t/2)\int_0^\infty dk\exp(-k^2t/2)(k^2xx_0)^{1+\mu}$,
and consequently by a simple scaling argument the long-time
expression for the probability distribution
\begin{equation}
\label{longbelow}
P(x,x_0,t)\simeq x x_0^{1+2\mu} e^{-|\epsilon|(x-x_0)}
e^{-\epsilon^2t/2}t^{-3/2-\mu}.
\end{equation}

The lifetime of a bubble of initial size $x_0$ created at $t=0$ follows
from Eq.~(\ref{longbelow}) by calculating the first passage time density
(FPTD) as time derivative of the survival probability, $W(t)=-\int_0^\infty
dx \partial P/\partial t$ \cite{vanKampen92}, or, via Eq.~(\ref{fokker}),
$W(t)=(1/2)[\partial P/\partial x+(2\mu/x-2\epsilon)P]_{x=0}$. This produces
\begin{eqnarray}
W(t)\simeq x_0^{1+2\mu}e^{|\epsilon|x_0}e^{-\epsilon^2t/2}t^{-3/2-\mu}.
\label{absorbbelow}
\end{eqnarray}
Below $T_{\text{m}}$, $\epsilon<0$ and the FPTD $W(t)$ decays exponentially.
The characteristic time scale is set by
\begin{eqnarray}
\tau=2/\epsilon^2\propto (T_{\text{m}}-T)^{-2},
\label{timescale}
\end{eqnarray}
that diverges as one approaches $T_{\text{m}}$. Finally, from
Eq.~(\ref{longbelow}) we infer that $P(x,t)\propto\exp(-c_1|\epsilon|(x+c_2
|\epsilon|t))$ with constants $c_i>0$, indicating that the profile of the
distribution has a drift $\sim|\epsilon|$ towards bubble closure at $x=0$.

At $T_{\mathrm{m}}$ ($\epsilon=0$) the FPTD falls off like a power law,
 $W(t)\propto t^{-\alpha}$ with scaling exponent
\begin{equation}
\label{exp}
\alpha=3/2+\mu.
\end{equation}
The parameter $\mu=c/2kT$ (with $\mu\approx 1$ at $T\approx T_{\text{r}}$)
is associated with the entropy loss of a closed polymer loop. Ignoring the
logarithmic entropic effects ($\mu=0$) we obtain $\alpha=3/2$, characteristic
of an unbiased random walk \cite{Hanke03}. From (\ref{absorbbelow}) we also
conclude that the mean bubble lifetime scales like
\begin{equation}
\label{mean}
\tau_{\text{mean}}\propto x_0/|\epsilon|\propto
x_0(T_{\text{m}}-T)^{-1},
\end{equation}
that diverges as the temperature is raised towards $T_{\text{m}}$.

\begin{figure}
\includegraphics[width=.64\hsize]{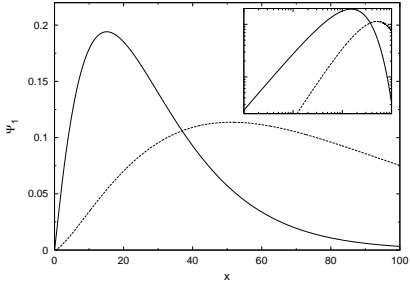}
\caption{Lowest bound state of the attractive well in $V(x)$ for
$T=10T_{ \mathrm{r}}$ (full line) and $T=2.2T_{\mathrm{r}}$
(dashed); with $T_{\mathrm{m}}=2T_{\mathrm{r}}$. Inset:
$\log$-$\log$ plot showing the power-law increase at $x=0$.}
\label{fig2}
\end{figure}

\emph{(ii) Long times for $T>T_{\text{m}}$.} Above $T_{\text{m}}$ ($\epsilon
>0$) the transition probability $P$ is controlled by the lowest bound states
in the attractive Coulomb potential. For the discrete spectrum we
have $E_n=\epsilon^2/2(1-(\mu/(\mu+n))^2)$, $n=1,2,\ldots$. The
lowest state is thus given by $E_1=\epsilon^2(\mu+1/2)/(\mu+1)^2$,
and the corresponding nodeless normalized bound state by
$\Psi_1(x)=Ax^{1+\mu}\exp{(-\mu\epsilon x/(\mu+1))}$ with
normalization constant
$A^2=[2\mu\epsilon/(\mu+1)]^{2\mu+3}/\Gamma(2\mu+3)$
\cite{Landau59c}. This bound state is localized at $\sim
1/(T-T_{\text{m}})$ and thus recedes to infinity as we approach the
melting temperature. In Fig.~\ref{fig2} we depict the lowest bound
state $\Psi_1$. From (\ref{prob}) we have $P(x,x_0,t)\sim
e^{\epsilon(x-x_0)}(x/x_0)^{-\mu} e^{-E_1t}\Psi_1(x)\Psi_1(x_0)$,
and we note that the dominant contribution to the distribution
originates from the region where the bound state peaks, i.e., at
$\sim 1/(T-T_{\text{m}})$. Inserting in (\ref{prob}) we obtain
\begin{eqnarray}
\nonumber
P(x,x_0,t)&=&A^2 x x_0^{1+2\mu}e^{(\epsilon/(1+\mu))(x-x_0(1+2\mu))}\\
&&\times e^{-\epsilon^2(1+2\mu)t/2(1+\mu)^2},
\label{longabove}
\end{eqnarray}
after some reduction. Note from (\ref{longabove}) that the profile of the
distribution drifts towards larger bubble sizes with velocity $\sim\epsilon$.
The associated FPTD becomes
\begin{eqnarray}
\nonumber
W(t)&=& A^2(1/2+\mu) x_0^{1+2\mu}e^{-\epsilon x_0(1+2\mu)/(1+\mu)}\\
&&\times e^{-\epsilon^2t(1+2\mu)/2(1+\mu)^2}.
\nonumber
\label{absorbabove}
\end{eqnarray}

\begin{figure}
\includegraphics[width=.64\hsize]{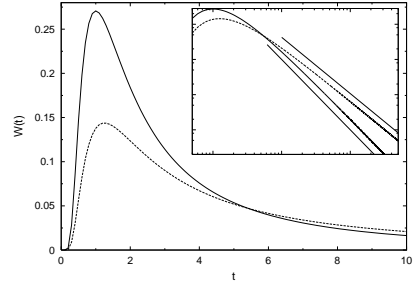}
\caption{Bubble lifetime distribution $W(t)$ from Eq.~(\ref{fptd})
for $T_m=2T_r$ (full line) and $T_m=10T_r$ (dashed). Inset:
$\log$-$\log$ plot of the power-law behavior at long $t$, with
slopes -2 and -1.6, as indicated by the straight lines.}
\label{fig3}
\end{figure}

\emph{Exact result at $T_{\text{m}}$.} At the melting temperature ($\epsilon
=0$) the bubble dynamics problem is equivalent to the case of a noisy
finite-time singularity studied in Ref. \cite{Fogedby02d}. The eigenstates of
$H$ are Bessel functions, $\Psi_k(x)=(kx)^{1/2}J_{1/2+\mu}(kx)$, and we obtain
after inserting in (\ref{prob}) the distribution
\begin{eqnarray}
\nonumber
P(x,x_0,t)&=&\frac{x^{1/2-\mu}}{x_0^{-1/2-\mu}} \int_0^\infty dk
e^{-k^2t/2}k^2J_{1/2+\mu}(kx)\\
&&\hspace*{2.6cm}\times J_{1/2+\mu}(kx_0),
\label{exact1}
\end{eqnarray}
or, by a well-known identity \cite{Lebedev72}, the explicit expression
\begin{eqnarray}
\nonumber
P(x,x_0,t)&=&\left(\frac{x}{x_0}\right)^{-\mu}(xx_0)^{1/2} t^{-1}
e^{-(x^2+x_0^2)/2t}\\
&&\times I_{1/2+\mu}(xx_0/t),
\label{exact2}
\end{eqnarray}
where $I_\nu$ is the Bessel function of imaginary argument
\cite{Lebedev72}. From (\ref{exact2}) we infer the FPTD
\begin{eqnarray}
\label{fptd}
W(t)=\frac{2x_0^{1+2\mu}}{\Gamma(1/2+\mu)}e^{-x_0^2/2t}(2t)^{-3/2-\mu},
\label{abs}
\end{eqnarray}
whose maximum at $t=x_0^2/(3+2\mu)$ assumes the value
\begin{eqnarray}
W_{\text{max}}=
\frac{2}{\Gamma(1/2+\mu)}\left(\frac{2e}{3+2\mu}\right)^{-3/2-\mu}
x_0^{-1/2}. \label{max}
\end{eqnarray}
In Fig.~\ref{fig3} we show the zero-size bubble distribution for two
different melting temperatures corresponding to different power-law
tails of the FPTD $W(t)$.

\emph{Correlations and finite size effects.} In typical experiments
measuring the fluorescence correlation of a tagged base-pair, DNA
bubble dynamics can be measured on the single molecule level
\cite{Altan-Bonnet03}. The correlation function $C(t)$ is
proportional to the integrated survival probability $C(t)\propto
\int_0^LP(x,x_0,t)dx$~, where $L$ is the chain length \cite{REM}.
From the definition of the FPTD we see that indeed $C(t)=1-\int_0^t
W(t)dt$. We find three cases:

(i) Below $T_{\text{m}}$ ($\epsilon<0$) we obtain from Eq.~(\ref{longbelow})
$C(t)=1-x_0^{1+2\mu}e^{|\epsilon|x_0}\int_0^te^{-\epsilon^2t'/2}(t')^{-3/2-\mu}
dt'$, or, in terms of the incomplete Gamma function $\gamma$ \cite{Lebedev72},
\begin{equation}
C(t)=1-x_0^{1+2\mu}e^{|\epsilon|x_0}(\epsilon^2/2)^{1/2+\mu}
\gamma(-1/2-\mu,\epsilon^2t/2).
\end{equation}
With $\gamma(\alpha,x)=\Gamma(\alpha)-x^{\alpha-1}e^{-x}$ for $x\rightarrow
\infty$ we obtain
\begin{eqnarray}
C(t)=\text{const.}+ x_0^{1+2\mu}\epsilon^{-2}e^{|\epsilon|x_0}t^{-3/2-\mu}
e^{-\epsilon^2 t/2}
\end{eqnarray}
for large $t$.  Note that the basic time scale of the correlations
is set by $\epsilon^{-2}\propto (T_{\text{m}}-T)^{-2}$. For
$t\ll\epsilon^{-2}$ the correlations show a power law behavior,
$C(t)\propto t^{-3/2-\mu}$; at long times $t\gg\epsilon^{-2}$ the
correlations fall off exponentially. The size of the time window
showing power law behavior increases as $T_m$ is approached. In
frequency space the structure function
$\tilde{C}(\omega)=\int\exp(i\omega t)C(t)dt$ has a Lorentzian line
shape for $|\omega|\ll\epsilon^2$, and power law tails for
$|\omega|\gg\epsilon^2$:
\begin{equation}
\tilde C(\omega)\sim\left\{\begin{array}{ll}
x_0^{1+2\mu} e^{|\epsilon|x_0}\left(\omega^2+(\epsilon^2/2)^2\right)^{-1},
& \mbox{for }|\omega|\ll \epsilon^2,\\
x_0^{1+2\mu} e^{|\epsilon|x_0}|\epsilon|^{-2}|\omega|^{1/2+\mu},
& \mbox{for } |\omega|\gg \epsilon^2.\end{array}\right.
\end{equation}

(ii) At $T_{\text{m}}$ ($\epsilon=0$) the exact expression for the
FPTD (\ref{abs}) combined with relation $C(t)=-\int_t Wdt$ yield
\begin{eqnarray}
C(t)=1-\frac{\Gamma(1/2+\mu,x_0^2/2t)}{\Gamma(1/2+\mu)}.
\end{eqnarray}
For short times $t\rightarrow 0$ the behavior
\begin{eqnarray}
C(t)=1-\frac{(x_0^2/2)^{\mu-1/2}}{\Gamma(1/2+\mu)}
t^{1/2-\mu}e^{-x_0^2/2t}
\end{eqnarray}
obtains, while in the long time limit $t\rightarrow\infty$,
\begin{eqnarray}
C(t)= \frac{2(x_0^2)^{1/2+\mu}}{(1+2\mu)\Gamma(1/2+\mu)}t^{-1/2-\mu}.
\end{eqnarray}

(iii) Above $T_{\text{m}}$ ($\epsilon>0$) the DNA chain eventually fully
denatures, and the correlations diverge in the thermodynamic limit. We can,
however, at long times estimate the size dependence for a  chain
of length $L$. From the general expression (\ref{prob}) we find
\begin{eqnarray}
C(t)\simeq e^{-\epsilon x_0} x_0^\mu\sum_n e^{-E_nt}\Psi_n(x_0)
\int_0^L e^{\epsilon x}x^{-\mu}\Psi_n(x)dx.~
\end{eqnarray}
With the lowest bound state $\Psi_1(x)=Ax^{1+\mu}\exp{(-\mu\epsilon x/
(\mu+1))}$ of energy $E_1=\epsilon^2(\mu+1/2)/(\mu+1)^2$, and after
integration over $x$, we obtain
\begin{eqnarray}
\nonumber
&&C(t)\propto A^2e^{-\epsilon x_0(2\mu+1)/(\mu+1)}
e^{-\epsilon^2((\mu+1/2)/(\mu+1)^2)t}x_0^{1+2\mu}\\
&&\times (1+\mu)\epsilon^{-2}
\left[1+(L\epsilon/(1+\mu)-1)e^{\epsilon L/(1+\mu)}]\right].
\end{eqnarray}
The correlations decay exponentially with the time constant $\sim
\epsilon^{-2}(\mu+1)^2/(2\mu+1)$. In frequency space the structure
function has a Lorentzian lineshape of width
$\sim\epsilon^2(2\mu+1)/(\mu+1)^2$, and for the size dependence one
obtains
\begin{equation}
C(t)\sim\left\{\begin{array}{ll}
L e^{\epsilon L/(1+\mu)}, & \mbox{for } \epsilon L/(1+\mu)\gg 1,\\
L\epsilon/(1+\mu), & \mbox{for } \epsilon L/(1+\mu)\ll 1
\end{array}\right..
\end{equation}
Note that close to $T_m$ the correlation function $C(t)\propto L$.

\emph{Summary and conclusion.} We demonstrated that the breathing
dynamics of thermally induced denaturation bubbles forming
spontaneously in double-stranded DNA can be mapped onto the
imaginary time Schr{\"o}dinger equation of the quantum Coulomb
problem. This mapping allows to calculate the distribution of bubble
lifetimes and the associated correlation functions, below, at, and
above the melting temperature of the DNA helix-coil transition.
Moreover, at the melting transition, the DNA bubble-breathing was
revealed to correspond to a one-dimensional finite time singularity.

Our analysis reveals non-trivial scaling of the first passage time density
quantifying the survival of a bubble after its original nucleation. The
associated critical exponents depend on the parameter $\mu$ stemming from
the entropy loss factor of the flexible bubble, and therefore on the ratio
$T_{\mathrm{r}}/T$ of reference and actual temperature. This correction
through $\mu$ decreases with increasing $T$. FPTD and correlations also
depend on the difference $T/T_m-1$, and therefore explicitly on the melting
temperature $T_{\mathrm{m}}$ (and thus the relative content of AT or GC
base-pairs). We also obtained the critical dependence of the characteristic
time scales of bubble survival and correlations on the difference $T-T_m$.
The finite size-dependence of the correlation function was recovered, as well.

The mapping of the of DNA-breathing onto the quantum Coulomb problem
provides a new way to investigate its physical properties, in
particular, in the range above the melting transition, $T>T_m$. The
detailed study of the DNA bubble breathing problem is of particular
interest as the bubble dynamics provides a test case for new
approaches in small scale statistical mechanical systems where the
fluctuations of DNA bubbles are accessible on the single molecule
level in real time.

Discussions with T. Ambj{\"o}rnsson, S. K. Banik, and A. Svane are
gratefully acknowledged. The present work has been supported by the
Danish Natural Science Research Council, the Natural Sciences and
Engineering Research Council (NSERC) of Canada, and the Canada
Research Chairs program.

%\bibliography{c:/user/manus/bib/bioarticles,c:/user/manus/bib/articles,c:/user/manus/bib/books}

\end{document}